\documentclass{elsart}

\usepackage{harvard}

\usepackage{graphics}
\usepackage{graphicx}
 \usepackage{psfig}

\def\url#1{{\ttfamily\def\/{/\discretionary{}{}{}}#1}}

\begin{document}

\begin{frontmatter}
\title{Evolution of Giant Radio Sources}


\author{C. H. Ishwara-Chandra\thanksref{ic}},
\thanks[ic]{E-mail: ishwar@ncra.tifr.res.in}
\author{D. J. Saikia\thanksref{djs}}
\thanks[djs]{E-mail: djs@ncra.tifr.res.in}

\address{National Centre for Radio Astrophysics, TIFR, Post Bag No. 3, 
Ganeshkhind, Pune - 411 007, India}

\begin{abstract}

We present radio images of two giant 
quasars from the Molonglo/1Jy sample, and make a comparative study of
giant radio sources selected from the literature with 3CR
radio sources of smaller sizes to investigate the evolution of giant sources,
and test their consistency with the unified scheme. 
The luminosity-size diagram shows that the giant sources
are less luminous than smaller-sized sources, consistent
with evolutionary scenarios where the giants have evolved from the
smaller sources, losing energy as they expand. For the giant sources
the equipartition magnetic fields are smaller, and inverse-Compton
losses with the microwave background radiation dominates over synchrotron
losses, while the reverse is true for the smaller sources.
The giant radio sources have core strengths similar to those of 
smaller sources of similar total luminosity; hence their large 
sizes are unlikely to be due to stronger nuclear activity. 
The radio properties of the giant radio galaxies and quasars are consistent
with the unified scheme.

\end{abstract}

\begin{keyword}
galaxies: active - quasars: general - radio continuum: galaxies
\end{keyword}

\end{frontmatter}

\section{Introduction}
\label{intro}

Giant radio sources (GRSs), defined to be those with a projected linear size
$\geq$1 Mpc (H$_\circ$=50 km s$^{-1}$ Mpc$^{-1}$ and q$_\circ$=0.5),
are the largest single objects in the Universe, and are extremely useful for
studying a number of astrophysical problems. These range from understanding
the evolution of radio sources and constraining orientation-dependent
unified schemes to probing the intergalactic medium at different redshifts
(Subrahmanyan \& Saripalli 1993, Ishwara-Chandra \&  Saikia 1999 for more 
detailed discussion).
In this paper we present radio images of two giant quasars 
from the Molonglo/1Jy sample, 0437$-$244 and 1025$-$229, and examine 
the evolution of GRSs and their consistency 
with the orientation-based unified schemes for radio galaxies and quasars.

\begin{figure}[t]
\hbox{
\hspace{0.4 in}
\psfig{figure=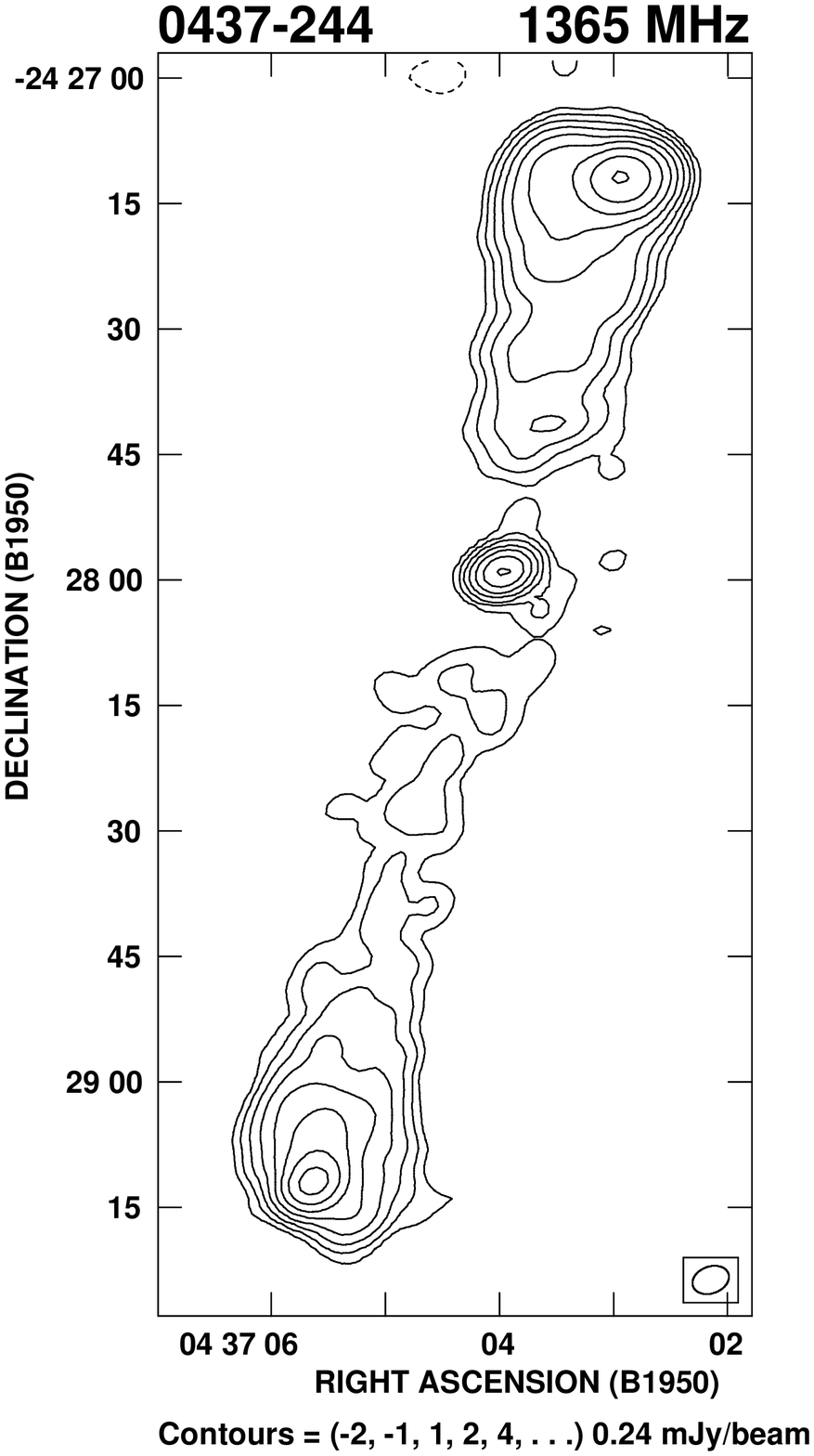,height=3.0in}
\hspace{0.75 in}
\psfig{figure=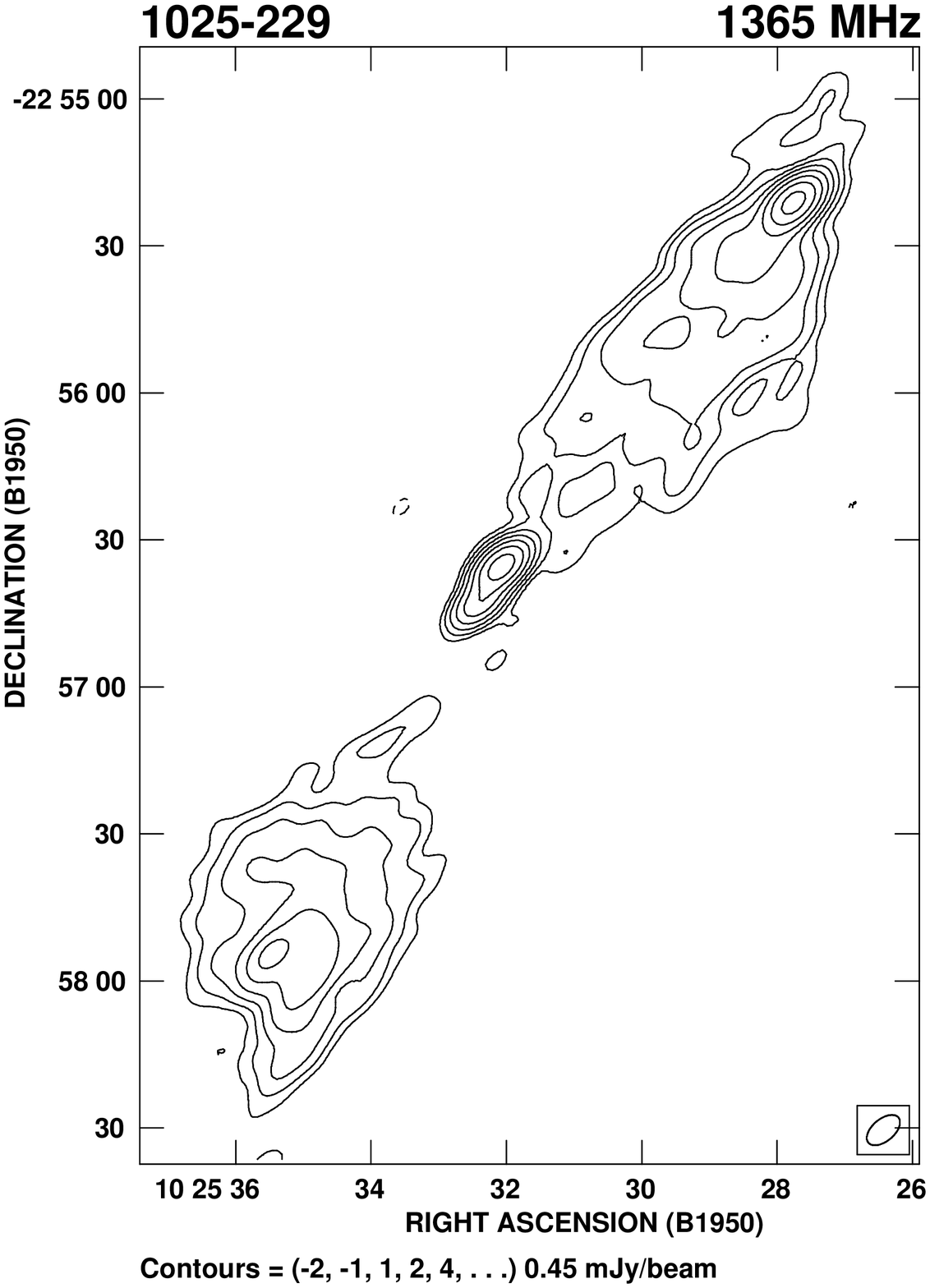,height=3.0in}
}
\caption{Radio images of the giant quasars at 1.4 GHz.}
\end{figure}

\section{Giant quasars from the Molonglo/1Jy sample}

\noindent {\bf 0437$-$244 :} This giant radio quasar (Fig. 1, left) is at a 
redshift of 0.84 and is at present the highest redshift giant quasar known 
and has a well defined FR II morphology. Its projected 
linear size is 1.06 Mpc, and has a core contributing about 10\% 
of the total flux density at an emitted frequency of 8\,GHz. 
The age estimates due to synchrotron radiative losses in the 
Kardashev-Pacholczyk model are 5.8$\times10^{7}$ and 
2.7$\times10^{7}$ yr for the northern and southern lobes respectively. 
The rotation measure between 1.4 and 5\,GHz are about 13.6$\pm$1.3 and 
4.2$\pm$3.4 rad m$^{-2}$ for northern and southern lobes respectively. The quasar shows 
no significant depolarization till about 1.4\,GHz. 

\noindent {\bf 1025$-$229 :} This is also a well-defined FR II radio source 
with two hotspots in the southern lobe, a core contributing about 12\%
of the total flux density of the source, and a possible jet-like 
structure close to the north of the radio core (Fig. 1, right). The redshift for this 
quasar is 0.309 and has a projected linear size of 1.11 Mpc. 
The spectral age estimates due to synchrotron
radiative losses are 7.8$\times10^{7}$ and 1.2$\times10^{8}$ yr for the
northern and southern lobes respectively. The rotation measures 
are $-$21.3$\pm$2.3, $-$15.3$\pm$0.9 and $-$21.5$\pm$3.6 rad m$^{-2}$ 
for the northern and southern lobes and the steep-spectrum jet-like 
feature close to the radio core respectively. Both the lobes do not show significant 
depolarization between 1.4 and 5\,GHz.

\begin{figure}[t]
\vspace{-1.05 in}
\hbox{
\hspace{1.0 in}
\psfig{figure=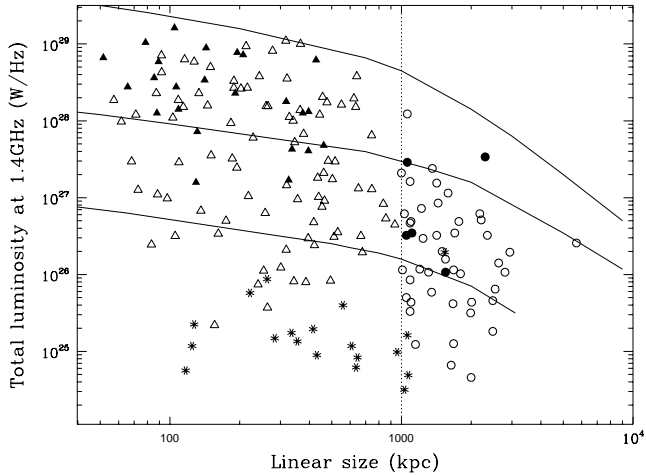,height=3.5in}
}
\caption{The luminosity - linear size or P-D diagram for all 3CR sources with
50 kpc $<$ D $<$ 1 Mpc and our sample of GRSs. The giant 
quasars and galaxies are shown by filled and open circles respectively;
while the 3CR quasars and galaxies are shown by filled and open
triangles respectively.  FR I sources have been marked with an asterisk.
The evolutionary scenarios for sources with jet powers
of 1.3$\times$10$^{40}$,  1.3$\times$10$^{39}$ and 1.3$\times$10$^{38}$ W from
Kaiser et al. (1997) are shown superimposed on the diagram.}
\end{figure}

\section{Evolution of giant radio sources} 

We investigate the evolution of the GRSs by plotting 
all known GRSs from the literature 
in a luminosity-linear size or P-D diagram along with the complete 
sample of 3CR radio sources (Laing, Riley \& Longair 1983) with sizes between 
50 kpc and 1 Mpc (Fig. 2). There is a clear deficit of GRSs
with high radio luminosity, suggesting that 
the luminosity of radio sources decrease as they evolve. 
We superimpose the evolutionary tracks suggested by Kaiser et al. (1997)
for three different jet powers in the P-D diagram and find that our 
sample of GRSs is roughly
consistent with their self-similar models where the lobes lose energy due to
expansion and radiative losses due to both inverse-Compton and 
synchrotron processes. In the models developed by Blundell et al.\,(1999)
the luminosity declines more rapidly than the Kaiser et al. tracks,
providing somewhat better fit to the upper envelope for 
large linear sizes. 
There is also a sharp cutoff in the sizes of 
the GRSs at about 3 Mpc, with only one exception, namely 3C236,
which has a size of 5.7 Mpc. 
To investigate whether there are 
larger sources which may have been missed, one requires low-frequency 
surveys with higher sensitivity to diffuse, low-brightness emission. 
Systematic searches for GRSs $>$ 3 Mpc using telescopes such as the GMRT 
would help clarify the late stages of their evolution.


We study the relative importance of synchrotron and
inverse-Compton losses in the evolution of GRSs
using the plot of linear size against the
ratio, B$_{eq}^2$/(B$_{ic}^2$ + B$_{eq}^2$), which represents the ratio
of the energy loss due to synchrotron radiation to the total energy loss due to
both inverse-Compton and synchrotron processes (Fig. 3).
It is clearly seen that synchrotron
losses dominate over inverse-Compton losses for almost all sources below
about a Mpc while the reverse is true for the GRSs.  
This also illustrates that inverse-Compton losses are 
likely to severely constrain the number of GRSs at high redshifts since 
the microwave background energy density increases as $(1 + z)^4$. 

\begin{figure}[t]
\vbox{
\vspace{-1.05 in}
\hspace{1.0 in}
\psfig{figure=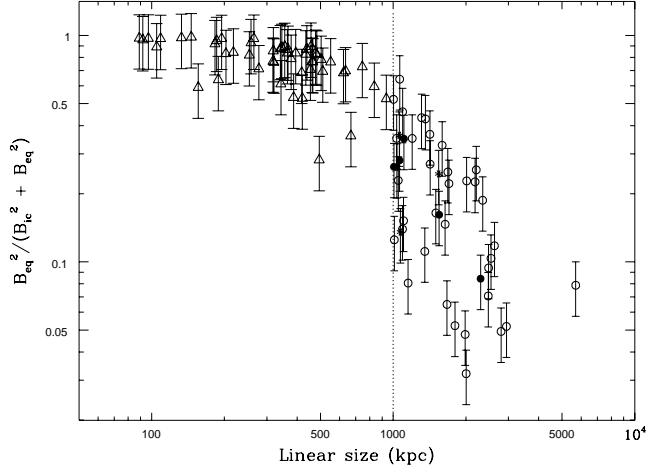,height=3.5in}
}
\caption{The projected linear size of the sample of giant and smaller sources
plotted against the ratio
 B$_{eq}^2$/(B$_{ic}^2$ + B$_{eq}^2$). 
The symbols are the same as in Fig. 2.}
\end{figure}

\section{Constraints on  Orientation  and environment} 

We have examined the suggestion that GRSs have powerful radio cores
than smaller sources (Gopal-Krishna, Wiita \& Saripalli 1989), which 
may be responsible for their large linear sizes. 
In Fig. 4, the fraction of the emission from the core at 
an emitted frequency of 8\,GHz is plotted against the total radio 
luminosity at 1.4\,GHz for GRSs as well as for smaller sources.
There is an inverse correlation between the degree of core prominence
and the total radio luminosity. However, GRSs have core strengths
similar to the smaller sources when matched in redshift or luminosity,
implying that GRSs are similar objects to the normal radio sources 
except for being larger and perhaps older. 

\begin{figure}[t]
\vspace{-1.05 in}
\hbox{
\hspace{1.0 in}
\psfig{figure=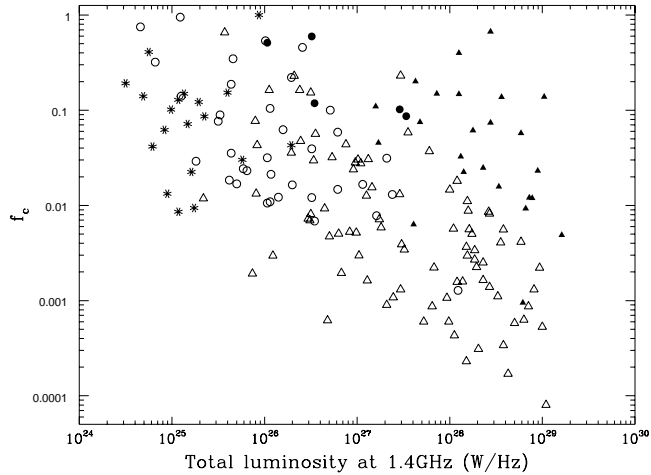,height=3.5in}
}
\caption{The total radio luminosity of the source plotted against the
fraction of emission from the core. The symbols are the same as in Fig. 2.}
\end{figure}

In order to check the consistency of the GRSs with the unified scheme for
radio galaxies and quasars, 
we have compared some of the orientation-dependent features 
such as core strength and core variability of giant radio galaxies 
and giant quasars. Although the available data are limited, 
these properties are consistent with the unified scheme.
We have also examined the environments around GRSs using their 
arm-length ratio and the misalignment angle. The arm-length ratio for 
GRSs are similar to those for the smaller sources, indicating that the
environment might be asymmetric on Mpc scales.  The 
misalignment angle for the GRSs is also similar to the smaller sources,
suggesting that their large sizes are unlikely to be due to a steadier 
ejection axis.

\end{document}